\documentclass[aps,pre,floatfix,twocolumn,superscriptaddress]{revtex4-2}
\usepackage[utf8]{inputenc}
\usepackage{amsmath}
\usepackage{amsfonts}
\usepackage{amssymb}
\usepackage{natbib}
\usepackage{graphicx}

\newcommand{\naybs}{NaYbS$_2$}
\newcommand{\nalus}{NaLuS$_2$}
\newcommand{\nayblus}{NaYb$_{1-x}$Lu$_{x}$S$_2$}

\newcommand{\llangle}{\langle\!\langle}
\newcommand{\rrangle}{\rangle\!\rangle}

\newcommand{\Nr}{N_{\rm rl}}

\usepackage{color}

\begin{document}


\title{
Diluting a triangular-lattice spin liquid: Synthesis and characterization of NaYb$_{1-x}$Lu$_{x}$S$_2$ single crystals
}
\author{Ellen H\"au{\ss}ler}
\affiliation{Fakult\"at f\"ur Chemie und Lebensmittelchemie, Technische Universit\"at Dresden, 01062 Dresden, Germany}
\author{J\"org Sichelschmidt}
\affiliation{Max Planck Institute for Chemical Physics of Solids, 01187 Dresden, Germany}
\author{Michael Baenitz}
\affiliation{Max Planck Institute for Chemical Physics of Solids, 01187 Dresden, Germany}
\author{Eric C. Andrade}
\affiliation{Instituto de F\'isica de S\~ao Carlos, Universidade de S\~ao Paulo,
C.P. 369, S\~ao Carlos, SP, 13560-970, Brazil}
\author{Matthias Vojta}
\affiliation{Institut f\"ur Theoretische Physik and W\"urzburg-Dresden Cluster
of Excellence ct.qmat, Technische Universit\"at Dresden, 01062 Dresden, Germany}
\author{Thomas Doert}
\affiliation{Fakult\"at f\"ur Chemie und Lebensmittelchemie, Technische Universit\"at Dresden, 01062 Dresden, Germany}

\date{\today}

\begin{abstract}
   Yb-based magnets, with a perfect triangular lattice of pseudospin-$1/2$ Yb$^{3+}$ ions, have emerged as candidates for realizing a quantum spin-liquid state, with NaYbS$_2$ being a prominent example. Here we present the solid-solution series NaYb$_{1-x}$Lu$_{x}$S$_2$ with well-defined single crystals over the entire substitution range $0\le x\le 1$. Chemical and structural analysis indicate a statistically homogeneous replacement of Yb$^{3+}$ by Lu$^{3+}$ ions. We magnetically characterize the relatively small single crystals using electron spin resonance (ESR). Below $30$\,K the ESR intensity can be well described by a Curie-Weiss function for all \textit{x}, with a decreasing Weiss temperature with increasing Lu content. This reduction of the average magnetic interaction upon Lu substitution is also supported by magnetization measurements. Importantly, no signs of magnetic or spin-glass order are detected down to $2$\,K for any $x$. For $x>0.5$ the ESR linewidth strongly increases, indicating the breakup of the magnetic system into disconnected clusters as expected from percolation physics.
   The experimental magnetization data are found to be in good agreement for all $x$ with results of classical Monte-Carlo simulations for a triangular-lattice Heisenberg model, amended with a small second-neighbor interaction.
   Taken together, our results establish NaYb$_{1-x}$Lu$_{x}$S$_2$ as a family of diluted triangular-lattice spin liquids.
\end{abstract}
\maketitle 


\section{Introduction}

Although the existence of quantum spin liquid (QSL) states was already proposed by Anderson nearly 50 years ago \cite{anderson_resonating_1973}, the number of comprehensively studied candidate materials has increased only recently \cite{field_guide}. A number of promising materials only qualify as proximate QSLs, such as $\alpha$-RuCl$_3$ \cite{banerjee_proximate_2016} and Na$_2$IrO$_3$ \cite{singh_antiferromagnetic_2010}, exhibiting long-range magnetic order at low temperatures. Others are debated because of considerable atomic disorder, such as YbMgGaO$_4$ \cite{li_rare-earth_2015,zhu_disorder-induced_2017} and H$_3$LiIr$_2$O$_6$ \cite{takagi18}, raising doubts on the intrinsic nature of the observed spin-liquid features. 

\begin{figure}[!t] 
\centering
\includegraphics[width=\columnwidth]{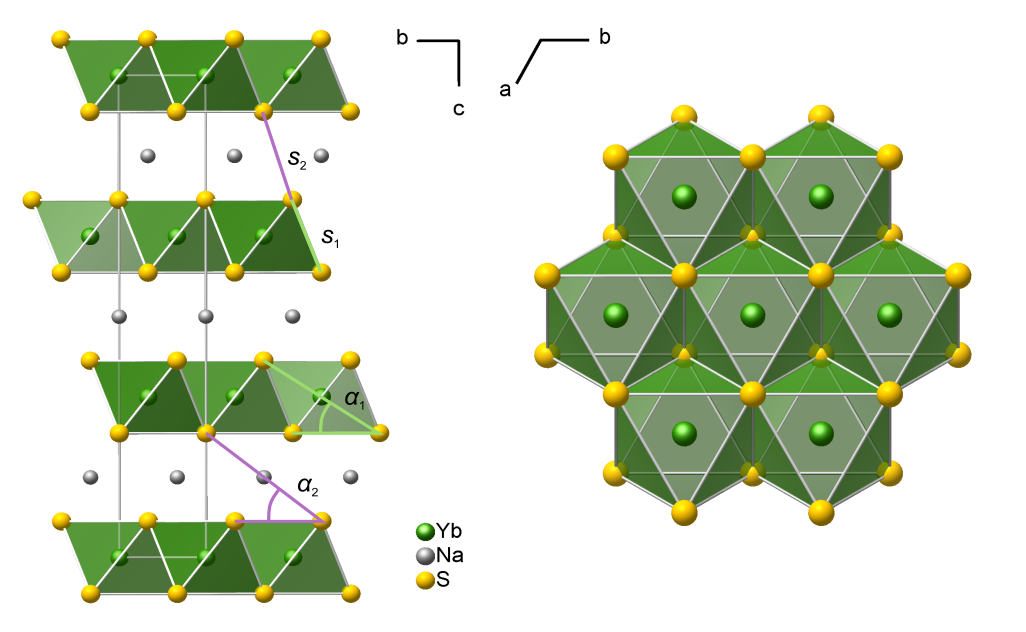}
\caption{Crystal structure of {\naybs} in $R\bar{3}m$ viewed along $a$ (left) and section of an YbS$_6$ layer (right). The unit cell is depicted in light grey. Interlayer sulfur - sulfur distances $s_1$ (light green) and $s_2$ (lavender) as well as the torsion angles $\alpha_1$ (light green) and $\alpha_2$ (lavender) of the YbS$_6$ and NaS$_6$ octahedra respectively are highlighted; see text for details.}
\label{fig:crystal structure}
\end{figure}

In the quest for new spin-liquid compounds, \textit{ABX}$_2$ compounds with $\alpha$-NaFeO$_2$ structures (with \textit{A} being a non-magnetic monovalent cation, \textit{B} a magnetic trivalent cation, and \textit{X} a divalent anion) attracted considerable attention since NaYbS$_2$  with Yb$^{3+}$ ions on a perfect triangular lattice was proposed as a putative QSL material \cite{baenitz_NaYbS2_2018,liu_rare-earth_2018}, Figure \ref{fig:crystal structure}. 
Further Yb-based compounds such as NaYbO$_2$ \cite{ranjith_field-induced_2019_180401,bordelon_field-tunable_2019} and NaYbSe$_2$ \cite{ranjith_anisotropic_2019_224417}  were investigated soon afterwards and also found to display no magnetic order down to 50 mK. Triangular lattice Yb compounds have thus established themselves as an important class of spin-orbit-entangled spin liquids \cite{schmidt_yb_2021}.
Note that the structure type of these compounds is frequently referred to as delafossite although the mineral delafossite, CuFeO$_2$, features \textit{A} cations in linear coordination \cite{shannon_1971}. In Na$REX_2$ and related compounds, \textit{A} and \textit{B} cations are both found in octahedral coordination. This atomic arrangement should be referred to as $\alpha$-NaFeO$_2$ type. $\alpha$-NaFeO$_2$ and CuFeO$_2$ are isopointal, they adopt the same space group type $R\bar3m$ with similar lattice parameters and respective atoms occupy the same Wyckoff sites but they are not isotypic in a strict sense. 
Indeed, \textit{ABX}$_2$ compounds with $\alpha$-NaFeO$_2$ type crystal structure provide a perfect geometrical basis for the realization of a QSL state, as different magnetic ions with predominant antiferromagnetic interactions can be placed on the \textit{B}-sites forming a regular triangular lattice. Atomic site disorder doesn't seems to play a dominant role, at least no evidence was found in single crystal diffraction data, NMR or ESR data of NaYbS$_2$ or NaYbSe$_2$, e.g. \cite{baenitz_NaYbS2_2018,ranjith_field-induced_2019_180401,ranjith_anisotropic_2019_224417,xing_field-induced_2019}. The layers of magnetically active \textit{B} ions are separated by non-magnetic \textit{AX}$_6$  octahedral layers, which gives the structure a quasi-two-dimensional character with respect to magnetism (Figure \ref{fig:crystal structure}). 

Theoretically, triangular-lattice spin models with either Heisenberg or spin-anisotropic exchange interactions have been shown to display QSL phases, accompanied by fractionalization and deconfinement, in certain ranges of their parameter space \cite{trumper93,imada14,zhuwhite15,iqbal16,chernyshev18,chernyshev19}. However, precise model parameters to describe the magnetism of such Yb compounds are not known to date.
With a clean candidate QSL material at hand, studying its behavior under chemical substitution defines a highly interesting field of research. This includes the physics of carrier doping to achieve metallic or superconducting states \cite{anderson_resonating_1973,rokhsar88,jiang21}, magnetic dilution \cite{nagaosa96}, and various forms of bond disorder \cite{zschocke15,bilitewski17,perkins19}. Magnetic dilution, achieved by replacing magnetic with non-magnetic ions, is particularly interesting, as the introduction of magnetic vacancies into quantum paramagnets is able to distinguish confined from deconfined phases \cite{sachdev_vojta01}; here ``(de)confined'' refers to the properties of the gauge field in a description in terms of fractionalized particles coupled to an emergent gauge field. In confined (i.e. conventional non-fractionalized) phases, vacancies introduce quantized magnetic moments which generically lead to magnetic \cite{nagaosa96} (or spin-glass \cite{andrade12}) order at low temperatures. A nice example is the quantum-dimer magnet TlCuCl$_3$ once small amounts of Cu are replaced by Mg \cite{fujisawa06}. In contrast, in deconfined (i.e. fractionalized) phases, no vacancy-induced moments are expected \cite{sachdev_vojta01,dommange03}, such that spin-liquid behavior survives dilution (although special counter-examples have been discussed theoretically \cite{sen15}). A number of theoretical studies have characterized spin liquids according to their response to the introduction of vacancies \cite{dommange03,willans10,sreejith16}. However, in-depth experimental studies on high-quality single crystals of corresponding dilution series of QSL candidates are rare.

Parenthetically, we note that bond disorder has been argued to induce spin-liquid-like behavior in magnets which display long-range order in the clean limit \cite{watanabe14a,li17b,yamaguchi17a,wu19c}. Moreover, strong dilution has been found to determine the magnetic state of the triangular-lattice system Y$_2$CuTiO$_6$ \cite{kundu20b}.

It is the purpose of this paper to study magnetic dilution for an Yb spin liquid on a triangular lattice. We investigate how the local and global magnetic properties evolve in the solid-solution series {\nayblus} with $ 0 \le x \le 1$ for which we are able to prepare single crystals covering the whole range of $x$. With increasing substitution of Yb$^{3+}$  by non-magnetic Lu$^{3+}$ we expect the network of exchange couplings to be weakened, such that the $x=0$ QSL eventually evolves into a state with single-ion paramagnetic properties as $x\to 1$.
We characterize the evolution and interplay of magnetic interactions in the substitution series {\nayblus} by electron spin resonance (ESR) and magnetic susceptibility measurements. Indeed, ESR is the method of choice for small single crystals \cite{luo18a}. For example, the nature of the ground state of a spin-liquid candidate in the organic salts could be successfully elucidated \cite{miksch21a}. {\naybs} has proven to be ideally suited for investigating the spin dynamics of Yb$^{3+}$  by ESR \cite{sichelschmidt19a}: a well-resolved and narrow Yb$^{3+}$  spin resonance allowed an accurate characterization of the \textit{g}-factor anisotropy and the static and dynamic properties of the Yb$^{3+}$ magnetic susceptibility.

Our experimental data are consistent with spin-liquid-like behavior of {\nayblus} in a wide composition range. We do not detect signs of magnetic order down to $2$\,K for all $x$. For $x>0.5$ we observe a strong increase in the ESR linewidth, indicating a breakup of the magnetic system into percolation clusters and suggesting that exchange interactions beyond nearest neighbors are small. Our measured temperature-dependent susceptibilities are in agreement with corresponding data from classical Monte Carlo simulations of a $J_1$-$J_2$ Heisenberg model with random dilution. Together, these results confirm that our crystals display a statistical distribution of Lu dopants and represent a family of diluted triangular-lattice quantum spin liquids.


\section{Experimental}

All samples of the solid-solution series {\nayblus} were prepared by a slightly modified procedure adopted from Masuda \emph{et al.} \cite{masuda_electrical_1999} starting from dried Na$_2$S (see \textit{Supplement} \cite{Supplement} for its preparation),  Yb$_2$O$_3$ (Heraeus, $99.9$\,\%), Lu$_2$O$_3$ (ChemPUR, $99.99$\,\%) and CS$_2$ (Honeywell, Riedel-de Haën, $99.9$\,\%) at 1050~°C in a protective argon atmosphere, following the idealized reaction equation:
\begin{align}
2\, \mathrm{Na}_2\mathrm{S} + 2\, RE_2\mathrm{O}_3 + 3\, \mathrm{CS}_2 \rightarrow 4\, \mathrm{Na}RE\mathrm{S}_2 + 3\, \mathrm{CO}_2
\label{eq:synth}.
\end{align}
Na$_2$S (2500 mg, 32 mmol, 29 equivalents) and the respective amounts of Yb$_2$O$_3$ \mbox{($1-x$ eqs.)} and Lu$_2$O$_3$ \mbox{($x$ eqs.)} were mixed in the molar ratios and filled in a glassy carbon boat which was  placed in a ceramic tube inside a tubular furnace (Figure S1, \cite{Supplement}). The large excess of Na$_2$S used in the reaction serves as flux material to foster crystal growth. The apparatus including a 1 L flask as CS$_2$ reservoir was flushed with argon for 30\,min before the furnace temperature was raised to 1050 °C within three hours under an unloaded stream of argon (ca. $2$\,L/h). While dwelling for one hour, CS$_2$ was transported to the reaction zone by bubbling argon ($\approx 5$\,L/h) through the reservoir.

Afterwards, the apparatus was cooled down under a stream of unloaded argon (ca. $2$\,L/h) to 600 °C within six hours before the furnace was switched off. Throughout the whole procedure, approximately 20~mL (about $330$\,mmol) of CS$_2$ were consumed.

The reaction product was leached with water to remove the flux material and the target compound was filtered-off and washed with water and ethanol. Leftovers of small black particles, presumably carbon, were decanted after ultrasonication in ethanol. After vacuum drying, the remaining product was found to consist of transparent, yellow hexagonal platelets with diameters ranging from approximately $10$ to $800$\,$\mu$m. Crystal surfaces show a faint tarnishing after exposure to moist air for several days or weeks. Crystals used for ESR and magnetization measurements were stored under inert conditions, thus. 

Phase purity and the evolution of the lattice parameters were evaluated by X-ray powder diffraction (PXRD). For this measurements, the samples were mixed with a Si standard and filled in glass capillaries. Rietveld refinements were done with TOPAS Academic \cite{coelho_topas_2018} in the range of 14 to 130° 2$\theta$, using the atomic positions of {\naybs} \cite{verheijen_flux_1975} as starting model. Single-crystal diffraction data of selected crystals were collected to confirm the Rietveld results. 

Elemental ratios were measured on several embedded and polished crystals via electron dispersive spectroscopy (EDX) in combination with imaging respective samples with a scanning electron microscope (SEM). The chemical composition of selected samples was additionally checked by optical emission spectroscopy (ICP-OES) and combustion analysis.
Optical band gaps were determined from diffuse reflection spectra using a Varian Cary 400 UV/Vis spectrometer. 
Magnetization measurements were performed with a Quantum Design superconducting quantum interference device (SQUID) vibrating sample (VSM) magnetometer. Crystals used for the magnetization measurements were taken from the same batches as those for the ESR experiments.
Electron spin resonance (ESR) experiments were performed at X-band frequencies ($\nu = 9.4$\,GHz) using a continuous-wave ESR spectrometer. The ESR experiments were performed on single crystalline platelets with a typical size of $0.4\times0.4\times0.05~\mathrm{mm}^3$. The sample temperature was set with a helium-flow cryostat allowing for temperatures between $2.7$ and $300$\,K \cite{campbell_continuous_1976}.

Further details on experimental methods are stated in the \textit{Supplement} \cite{Supplement}.


\section{Results and discussion}

\subsection{Crystal growth, optical and chemical properties}

\begin{figure}[ht] 
\centering
\includegraphics[width=\columnwidth]{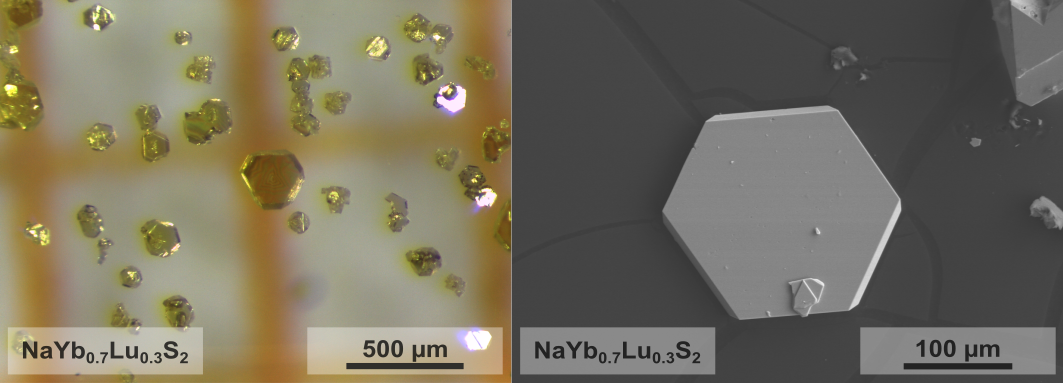}
\caption{Images of crystals taken under an optical microscope (left panel) and by scanning electron microscopy (right panel).}
\label{fig:crystal images}
\end{figure}

{\nayblus} crystals grow as hexagonal platelets or antiprisms with an aspect ratio of \mbox{(10 - 20):1}. The typical morphology of the crystals can be seen in the images in Figure \ref{fig:crystal images} taken with a light microscope and a Scanning Electron Microscope (SEM), respectively. Over the major part of the substitution range, the crystals remain bright yellow, like the ternary border compound {\naybs}. Crystals with high lutetium content \mbox{($x > 0.8)$} are pale yellow.

The band gaps of {\naybs} and {\nalus} were determined to $2.74$ and $3.94$~eV, matching the reported data of $2.7$~eV \cite{liu_rare-earth_2018} and $4.08$~eV \cite{jary_optical_2015}, respectively. The band gaps of {\nayblus} remain constant at around $2.73$\,eV until $x \le 0.8$, but jump to $3.87$\,eV for $x = 0.9$ (Figure S2 \cite{Supplement}). 
According to band structure calculations of isostructural rare-earth-metal based compounds the band gap mainly results from the distance between the occupied chalcogenide \textit{p}-states and the \textit{RE\,d}-state dominated conduction band \cite{deng_new_2002}. The sudden step in the evolution of the band gaps and the corresponding color change can't be explained by the observed smooth change of the lattice parameters, see below. The UV/vis spectrum of {\nalus} shows only one absorption edge at 315~nm, whereas the spectrum of {\naybs} exhibits an additional broad band at 360~nm (see Figure S3 \cite{Supplement}). The latter is attributed to a sulfur to $RE$ charge transfer (CT) transition \cite{havlak11} and can be observed in all compounds with $x \le$ 0.8. As the low-energy slope of this CT transition is now used to determine the absorption edge, a considerable drop in band gaps is computed in accordance with the colors of the samples.  

Upon grinding of the crystals under ambient conditions, a faint H$_2$S odour was recognized. X-ray powder diffractograms of ground samples showed small reflections of the respective oxysulfides $RE_2$O$_2$S, so that we assume the following slow decomposition of the ground sample fostered by humidity from air and/or adsorbed water:
\begin{align}
2\, \mathrm{Na}RE\mathrm{S}_2 + 4\, \mathrm{H}_2\mathrm{O} \rightarrow RE_2\mathrm{O}_2\mathrm{S} + 2\, \mathrm{NaOH} + 3\, \mathrm{H}_2\mathrm{S}.
\label{eq:decomp}
\end{align}
Reflections of NaOH were not observed in PXRD of the ground samples, probably due to the formation of non-crystalline hydrates of NaOH and/or Na$_2$S. Traces of the oxysulfide were only found after grinding, i.e. in samples with small particles and large surfaces. Crystals stored under inert conditions are stable for years. 


\subsection{Chemical and structural characterization}

Chemical analyses confirm the composition of the samples within the error of the respective method, Table S~I \cite{Supplement}. The EDX-based sulfur contents of the crystals \mbox{($47(1)$\,at.-\%)} was always found slightly below the bulk value from ICP-OES and combustion analysis \mbox{($49.1(8)$\,at.-\%)}; we attribute this to the embedding and wet-chemical polishing procedure applied before EDX. Carbon \mbox{($< 1$\,at.-\%)} and oxygen amounts \mbox{($< 0.6$\,at.-\%)} of the bulk polycrystalline samples were generally considered as insignificant for the ESR and magnetic measurements, which were all performed on selected single crystals. \\
No peak splitting or additional reflections were found in the X-ray powder diagrams of the samples. Instead, all reflections can be indexed with space group $R\bar3m$ and unit cell dimensions close to the border phases {\naybs} and {\nalus} so that a complete substitution series {\nayblus} ($0\le x \le 1$) without phase separation or symmetry reduction can be assumed. The PXRDs of all samples exhibit a strong preferred orientation along the [001] direction. The lattice parameters were refined in the course of the Rietveld fits using the fundamental parameter approach and a manually modelled background to take care of the enhanced background at low angles (between 14° and 30° $2\theta$) caused by the glass capillary.

\begin{figure}[!ht] 
\centering
\includegraphics[width=\columnwidth]{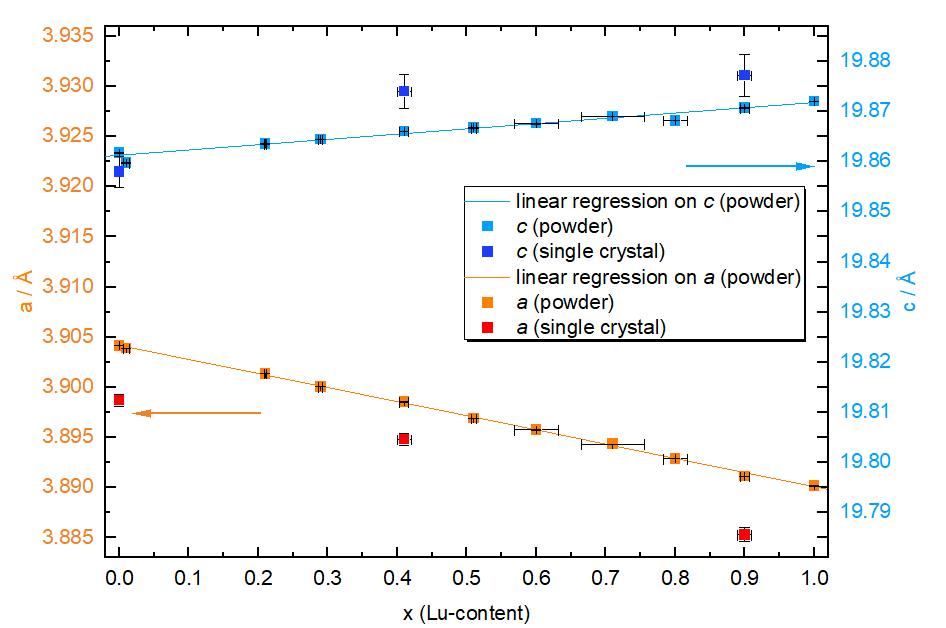}
\caption{Refined lattice parameters $a$ and $c$ of {\nayblus} samples derived from powder and single crystal data.}
\label{fig:lattpar_vs_x}
\end{figure}

The refined lattice parameters are stated in Table S~II \cite{Supplement} and graphically depicted in Figure\,\ref{fig:lattpar_vs_x} against the amount of Lutetium. For both ternary boundary phases {\naybs} $(x = 0)$ and {\nalus} $(x = 1)$, literature structure data were confirmed. The atomic positions of {\naybs} were used as start model for the Rietveld refinements, considering a mixed occupancy of the Wyckoff site 2\textit{a} with Yb and Lu for all compounds with $0<x<1$. As the electron count of the two rare-earth elements differ by only one, a free refinement of the occupation parameters is merely possible. Instead, the site occupancy factors for both elements were taken from the analytical data (EDX, ICP-OES) and kept fix during the refinements. We found neither evidence for other mixed occupied sites or site defects nor indications for an ordered superstructure at any degree of substitution from these fits. The result of the Rietveld refinement for NaYb$_{0.71}$Lu$_{0.21}$S$_2$ is depicted as one example in Figure S4 \cite{Supplement}.

No peak spitting was observed in the diffraction images of selected single crystals of {\nayblus} (with $x = 0,\,0.4\, \mbox{and } 0.9$) samples either, confirming the PXRD based Rietveld results. The respective structure models could be refined to good agreement values in space group $R\bar{3}m$; no additional peak positions or suspicious displacement parameters or residual electron density maxima were found. The refinement of the Na and $RE$ site occupancy factors did not deviate from unity within standard deviations, i.e. there is no evidence for Na/$RE$ site disorder from single crystal diffraction data. Figure \ref{fig:crystal structure} depicts the crystal structure of {\naybs} as an example. Crystallographic data and refinement parameters are compiled in Tables S~III, S~IV and S~V, respectively \cite{Supplement}. 

As can be seen from Figure~\ref{fig:lattpar_vs_x}, the lattice parameter $a$ decreases linearly with increasing Lu content. This can be easily understood as the larger Yb$^{3+}$ ion ($r = 1.008\, \mathring{\mathrm{A}} $) is progressively replaced by the smaller Lu$^{3+}$ ion ($ r = 1.001\, \mathring{\mathrm{A}} $; both radii are given for coordination number 6 \cite{shannon_revised_1976}). The $c$ lattice parameter, however, gets elongated with increasing Lu content. Note that the change of the lattice parameter $c$ is about seven times smaller than the change in $a$ (0.05~\% for $c$, 0.35~\% for $a$), so that the unit cell volume decreases slightly (Figure S5, \cite{Supplement}). The slight deviations between powder and single crystal data can be attributed to the use of different devices and wavelengths.

Upon substitution not only the lattice in the $ab$ plane is compressed but also the YbS$_{6}$ octahedra show a similar trend, as can be seen from the decreasing interlayer sulfur-sulfur distances $s_1$ and the torsion angle $\alpha_1$ of YbS$_{6}$ octahedra (Figure \ref{fig:crystal structure}). Unfortunately, the powder-based structure data show some scattering, although displacement parameters and site occupation factor were kept fix to reduce correlations in the Rietveld refinements. The single-crystal data, however, indicate the same trend. The overall changes of 0.6\,\% for $s_1$ and 0.2\,\% for $\alpha_1$ are in the same range as those of the lattice parameter $a$. Consistent with the slight increase in lattice parameter $c$ the NaS$_{6}$ octahedra slightly increase as well, which can be concluded from the sulfur - sulfur interlayer distances $s_2$ and the torsion angle $\alpha_2$. The changes of the interatomic distances and torsion angles are depicted in Figures S6 and S7 \cite{Supplement}.


\subsection{ESR spectroscopy}

\begin{figure}[tb] 
\centering
\includegraphics[width=\columnwidth]{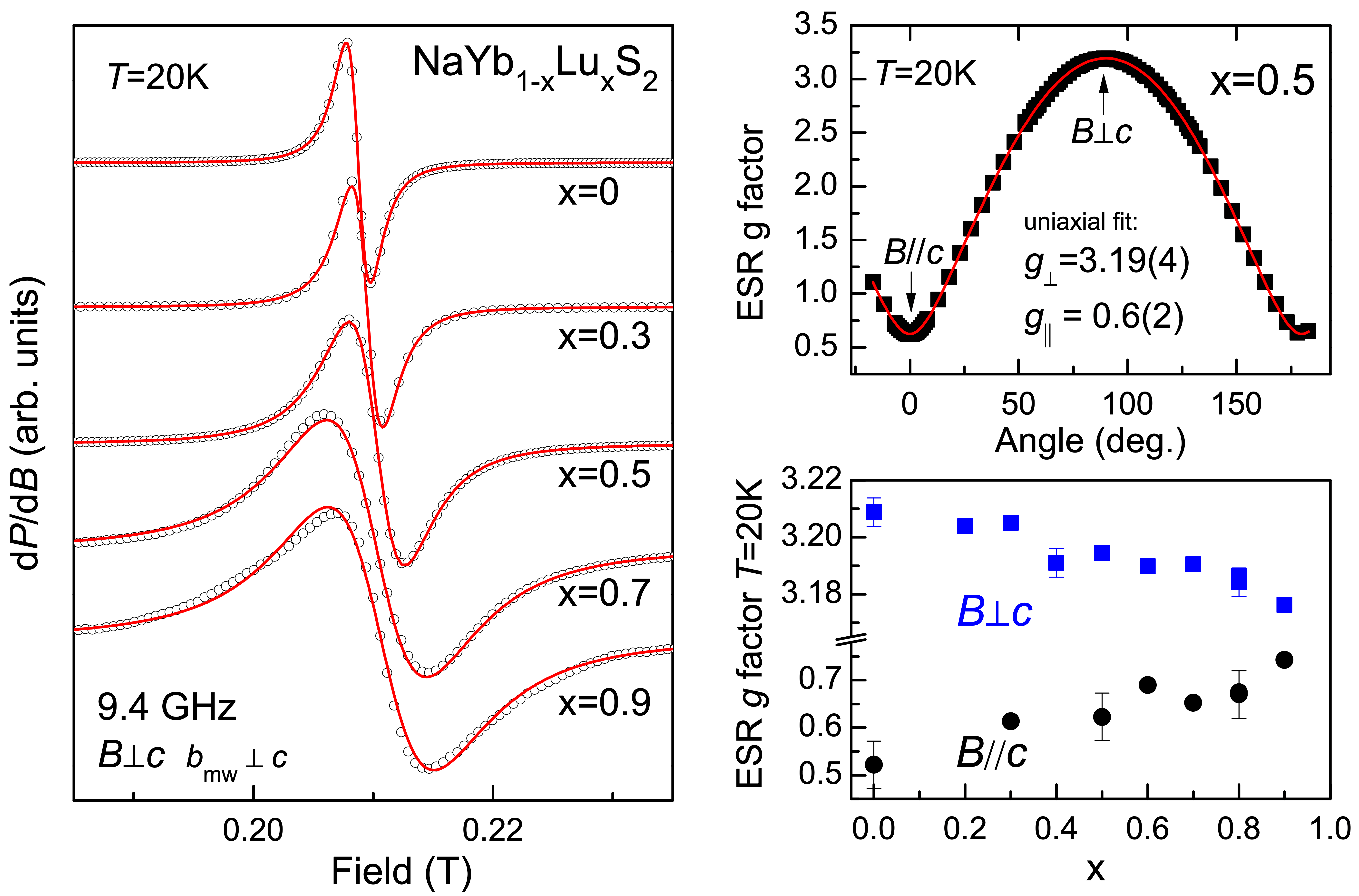}
\caption{Left frame: Typical X-band ESR spectra at $T = 20$\,K of Yb$^{3+}$ for selected $x$ in {\nayblus} single crystals and for indicated field orientations. Solid lines indicate single Lorentzian line shapes. Upper right frame: ESR $g$ factor dependence for $x = 0.5$, depending the field orientation with respect to the crystalline $c$ axis. Lower right frame: Yb amount dependence of the $g$ factor anisotropy for $T = 20$\,K.}
\label{fig:ESRFig1}
\end{figure}

Typical ESR spectra are shown in Figure \ref{fig:ESRFig1}, left frame. The deviations from an ideal Lorentzian shape are small but visible. For samples with $x\ge0.8$ a multiple line structure appears which is consistent with a Yb-hyperfine split resonance. As shown in Figure S8 \cite{Supplement} this hyperfine split structure is most pronounced for $x=0.9$ which indicates a very weak interaction among the Yb spins for this concentration. 
Delafossite-related Yb compounds show strong anisotropies of their Yb$^{3+}$ resonance field \cite{sichelschmidt_effective_2020} caused by uniaxial $g$ value anisotropies. For {\nayblus} this anisotropy is retained for all investigated Yb-contents and is exemplarily shown for $x = 0.5$ in Figure \ref{fig:ESRFig1} (upper right frame). A clear $x$ dependence of the anisotropy $g_{\perp}/g_{\parallel}$ can be resolved as shown in the lower right frame of Figure \ref{fig:ESRFig1}.

\begin{figure}[tb] 
\centering
\includegraphics[width=\columnwidth]{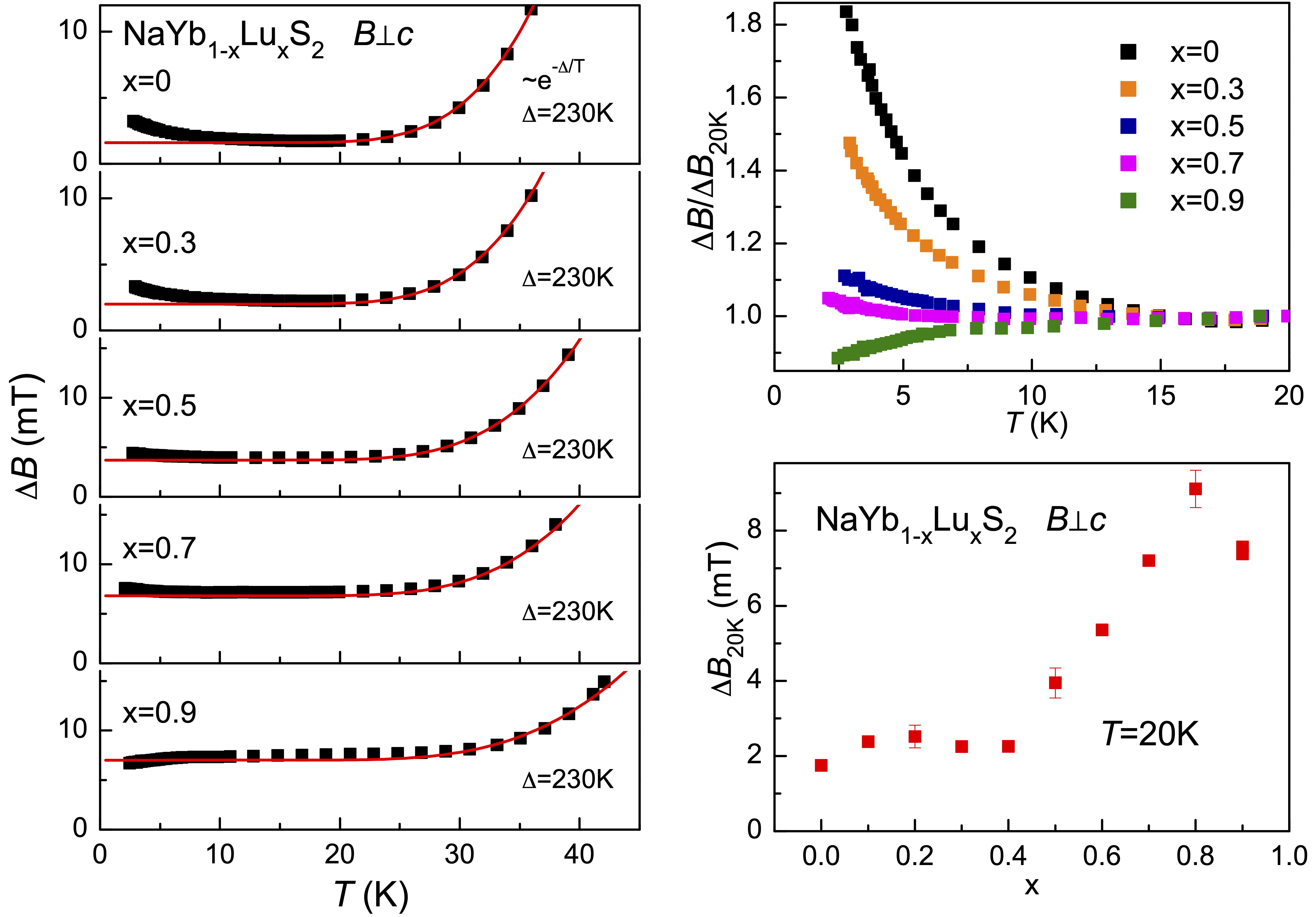}
\caption{Temperature- and $x$-dependence of the ESR linewidth $\Delta B$ of {\nayblus} single crystals. Left frame: Towards higher temperatures the solid lines indicate a relaxation via the first excited CEF level of Yb$^{3+}$ at $\Delta$K. Right upper frame: Low-temperature behavior of $\Delta B$ normalized by its value at $T = 20$\,K. Right lower frame: $x$-dependence of the linewidth at $T=20$\,K.}
\label{fig:ESRFig2}
\end{figure}

The ESR linewidth behavior $\Delta B(T)$ is shown in Figure \ref{fig:ESRFig2}. Towards high temperatures, as seen in related Yb compounds \cite{sichelschmidt_effective_2020}, $\Delta B(T)$ shows the same exponential increase being characteristic for an Orbach-type spin lattice relaxation $\Delta B\propto 1/[\exp(\Delta /T)-1]$ (red solid lines). This relaxation involves the first crystalline-electric field (CEF) split electronic energy level $\Delta$ above the ground state \cite{abragam_electron_1970}. We obtained $\Delta~=~230~\pm~50~\text{K}$, independent on $x$, indicating that mainly properties of the CEF ground state are affected by changes of the Yb$^{3+}$ amount, i.e., changes in the CEF distortion. Noteworthy, changes of the ground-state properties are well resolved in changes of the $g$-factor anisotropy, see Figure \ref{fig:ESRFig1}.

Towards low temperatures, as illustrated by a normalized $\Delta B(T)$ representation, Figure \ref{fig:ESRFig2} right upper frame, the linewidth behavior clearly depends on the Yb$^{3+}$ amount (see also the non-normalized plot Figure S9, \cite{Supplement}). For $x < 0.8$ the linewidths display a moderate broadening, indicating the increasing influence of inter-site spin correlations \cite{sichelschmidt_effective_2020}. For $x \ge 0.8$ $\Delta B(T)$ shows, instead of an increase, a decrease becoming pronounced below $T~\approx ~10$\,K. Such temperature dependence is unveiled by a diminished spin-spin correlation as it appears concomitant to a single-ion typical hyperfine structure in the line shape. It indicates a direct relaxation of spin-orbit coupled spins to lattice vibrations through a modulation of the ligand field, which leads to a linear temperature dependence if a direct phonon process dominates, for instance \cite{abragam_electron_1970,wolfe_paramagnetic_1971}.

The linewidth also clearly depends on the Yb content as illustrated for $T = 20$\,K in the right lower frame of Figure \ref{fig:ESRFig2}. With decreasing Yb-content $\Delta B$ first stays at values around $2$\,mT for $x < 0.5$ before rising up to $\Delta B$ = 7.5 mT for $x=0.9$. This indicates for $x > 0.5$ an additional broadening arising from an inhomogeneous Yb-distribution as found for disconnected magnetic clusters, for instance. Note that at the same time the above mentioned contribution of inter-site spin correlations is strongly reduced but still visible up to $x \le 0.8$ in the line broadening towards low temperatures.

\begin{figure}[ht] 
\centering
\includegraphics[width=\columnwidth]{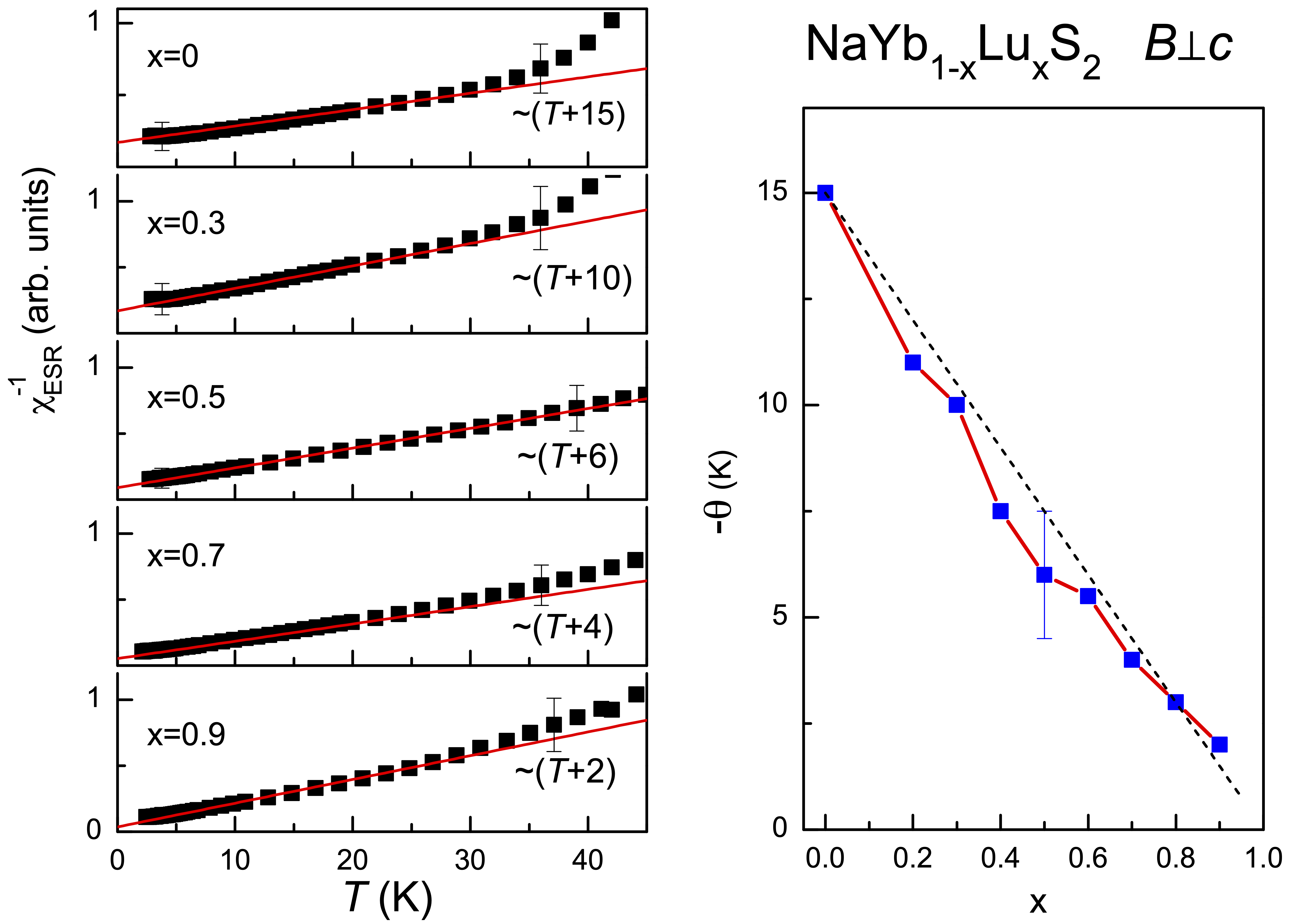}
\caption{Temperature dependence of the reciprocal ESR intensity $\chi_{\rm ESR}^{-1}$ for various $x$ in {\nayblus} single crystals with their $c$-axes aligned perpendicular to the external field $B$ and the microwave field $b_{mw}$. Solid lines indicate a low-temperature linear behavior $\propto(T-\theta)$ with Weiss temperatures $\theta$ as plotted in the right frame. Dashed line corresponds to $(1-x)15$\,K.}
\label{fig:ESRFig3}
\end{figure}

\begin{figure}[h] 
\centering
\includegraphics[width=\columnwidth]{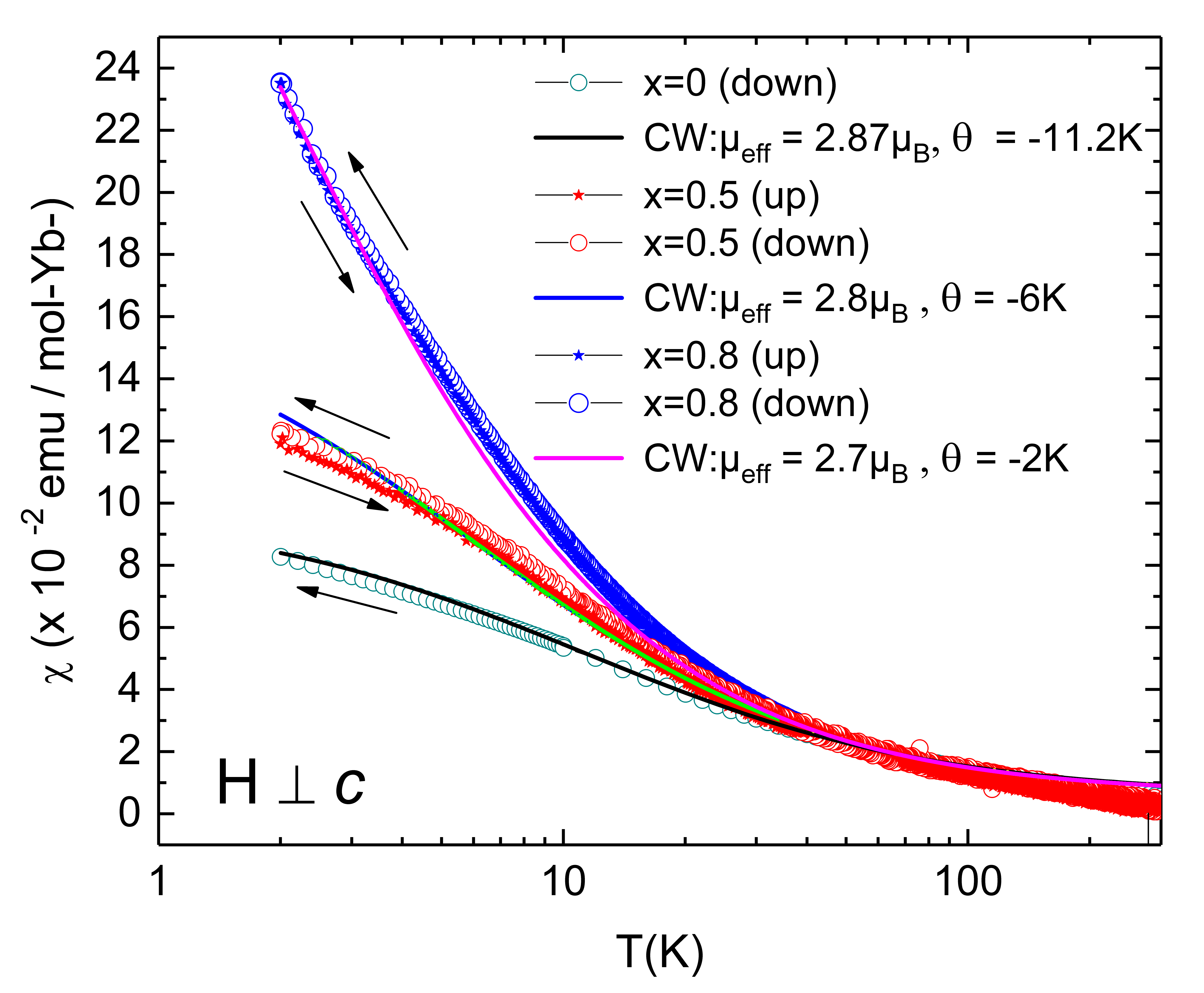}
\caption{
Magnetic susceptibility for {\nayblus} cystals with $x = 0,\,0.5\, \mbox{ and } 0.8$ for fields applied perpendicular to the crystallographic $c$ direction. The solid lines represent a Curie-Weiss (CW) fit to the data below $80$\,K. Measurements for $x = \,0.5\, \mbox{ and } 0.8$ are performed in zfc(up)/fc(down) mode. 
}
\label{fig:Lu_substitution_NaYbS2}
\end{figure}

The ESR intensity $I_{\mathrm{ESR}}\equiv \chi_{\text{ESR}}$ is a direct measure of the spin probe magnetic susceptibility along the direction of the microwave magnetic field $b_{mw}$ \cite{gruner_anisotropic_2010}. As shown in Figure \ref{fig:ESRFig3}, below approximately $30$\,K, $\chi_{\text{ESR}}$ is consistent with a Curie-Weiss law $\propto (T-\theta)^{-1}$ with negative Weiss temperatures $\theta$. A decreasing Yb$^{3+}$ content leads to a continuously decreasing $-\theta$ (right frame in Figure \ref{fig:ESRFig3}) as expected for decreasing antiferromagnetic correlations among the Yb$^{3+}$ spins. This behavior agrees well with measurements of the bulk magnetic susceptibility as illustrated for $x = 0, 0.5, \mbox{and } 0.8$ in Figure \ref{fig:Lu_substitution_NaYbS2}.
Note, however, that the Weiss temperatures obtained from bulk susceptibility and ESR intensity do not exactly match. The reason for such difference, being also observed in other compounds \cite{schmidt_yb_2021,gruner_anisotropic_2010,arjun19a,baenitz21a}, could be that spin correlations are probed by ESR locally at the Yb-sites in contrast to magnetic susceptibility which is a non-local and non-dynamical probe.

\subsection{Susceptibility}
Figure 7 shows the temperature dependence of the magnetic susceptibility of {\nayblus} crystals with $x = 0,\,0.5\, \mbox{and } 0.8$ studied at the ESR magnetic field of about $0.21$\,T for fields applied perpendicular to the crystallographic $c$ direction. Below $80$\,K the data can be well described by a Curie-Weiss curve. The Curie-Weiss fits show that the Weiss temperature decreases sharply with increasing Lu amount while the effective moment at Yb hardly changes. These results are in a good agreement with the ESR measurements, i.e., a strong change in the Weiss temperature is observed  while the $g$ factor decreases only moderately.
The zero-field-cooled (zfc) and field-cooled (fc) mode measurements do not show any branching or pronounced hysteretic behavior. We take this as evidence for the absence of magnetic domains, structural clusters, or glassy-like effects. As shown for the example of {\nayblus} with $x = 0.8$, only a small difference in the order of about $0.01$\,T, is visible in the magnetization curve (at $10$\,K) with ascending and descending magnetic field (Figs. S10 and S11, \cite{Supplement}).

\subsection{Monte-Carlo simulations}

A minimal model to capture the physics of the delafossite-like spin-liquid candidate compounds is the $J_1$-$J_2$ Heisenberg model on the triangular lattice,
\begin{align}
\mathcal{H} & =
J_1 \sum_{\langle ij\rangle} \vec{S}_{i}\cdot\vec{S}_{j} +
J_2\sum_{\llangle ij \rrangle}\vec{S}_{i}\cdot\vec{S}_{j}
\label{eq:j1j2}
\end{align}
with antiferromagnetic first-neighbor and second-neighbor couplings $J_1$ and $J_2$, respectively. The classical ground state of this model depends on the ratio $J_2/J_1$: For $J_2/J_1<1/8$ a coplanar 120$^\circ$ state is realized, while $J_2/J_1>1/8$ yields a collinear four-sublattice state \cite{chubukov92}. For quantum spins $S=1/2$, numerics indicates the existence of a fractionalized quantum spin liquid for $0.08<J_2/J_1<0.16$ whose precise nature, however, has not been clarified beyond doubt \cite{trumper93,imada14,zhuwhite15,iqbal16}. With spin anisotropies included, the nearest-neighbor exchange model has also been shown to host quantum spin-liquid phases \cite{chernyshev18,chernyshev19}. At this point there is no definite consensus on the microscopic model faithfully describing the triangular spin-liquid candidates.

To theoretically assess the effect of dilution on the magnetic properties of NaYbS$_2$ at elevated temperatures, we confine ourselves to a study of \eqref{eq:j1j2} in the classical limit. We perform classical MC simulations on lattices of linear size $L$ and periodic boundary conditions. Our spins are then replaced by classical vectors of fixed length $S$. Depletion is simulated by randomly removing a fraction $x$ of spins, with the total number of spins $N_{\mathrm{s}}=\left(1-x\right) N$. We perform equilibrium MC simulations using single-site updates with a combination of the heat-bath and microcanonical (or over-relaxation) methods. For simulations at high temperatures, the simulations reach equilibrium quickly and tens of thousands of MC steps per spin are sufficient to evaluate the thermal averages. Disorder averages are taken over $\Nr$ samples, with $\Nr\sim100$.

Before discussing numerical results, we recall that previous work on vacancies in non-collinear triangular-lattice antiferromagnets has shown that an isolated vacancy induces a uniform magnetic moment in the ground state \cite{wollny11}. As a result, a weak residual Curie term in the susceptibility at very low $T$ is expected even for small nonzero $x$ \cite{wollny11,zhitomirsky15}. For a model with nearest-neighbor coupling only, the threshold for site percolation is $x_p=1/2$, such that a system with statistically distributed vacancies and $x>x_p$ consists of magnetically disconnected finite-size clusters only. Consequently, larger Curie contributions (per spin) are expected, arising from clusters with an odd number of sites, which grow as $x\to1$. Small non-zero values of $J_2$ will produce only small corrections to this picture.

\begin{figure}[ht] 
\centering
\includegraphics[width=\columnwidth]{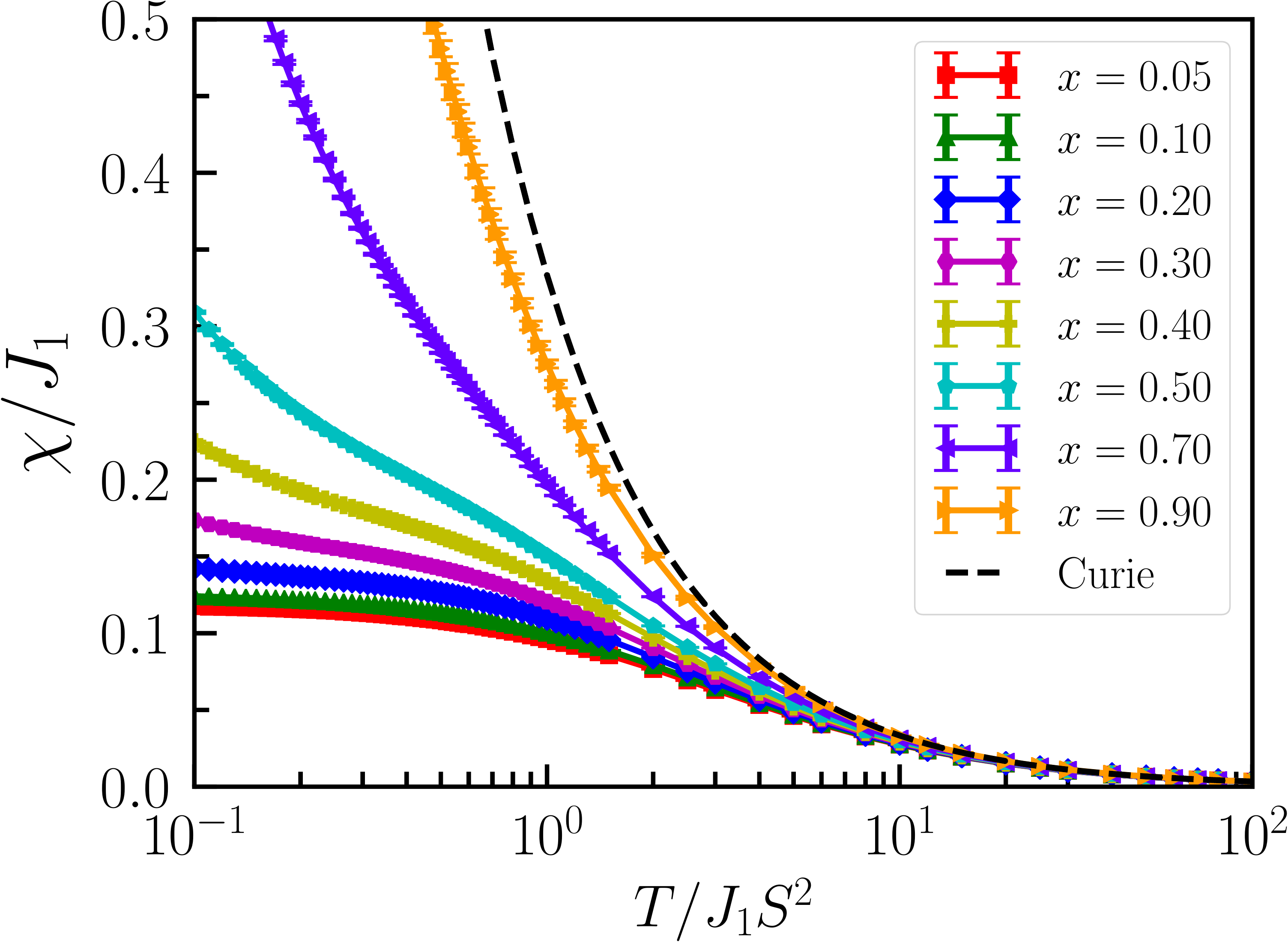}
\caption{
Simulation results for the magnetic susceptibility per spin, $\chi(T)/N_s$, of the classical Heisenberg model \eqref{eq:j1j2} with $J_2/J_1=0.12$ for different concentrations $x$ of magnetic vacancies. The simulations were performed for $L=30$; finite-size effects are minimal for the temperatures shown.
}
\label{fig:mc_chi}
\end{figure}

Simulation results for the normalized linear-response magnetic susceptibility $\chi(T)/N_s$ are shown in Figure~\ref{fig:mc_chi} for different levels of dilution. While $\chi(T)$ follows a Curie-Weiss law for all $x$ at elevated $T$, it tends to saturate upon cooling for small $x$ in the $T$ range shown, but strongly increases for large $x$. In particular, the $x=0.9$ result is not far from a pure Curie law, consistent with the above considerations.

The Curie-Weiss law can be analyzed quantitatively: High-temperature expansion for the diluted Heisenberg antiferromagnet yields the exact result \cite{binder86} $k_B|\theta| = (1-x) S(S+1) z (J_1+J_2)/3$, with $z=6$ being the number of nearest-neighbor sites. This qualitative $x$ dependence, $\theta(x) = (1-x) \theta(x\!=\!0)$, is in good agreement with the ESR-extracted linearity in Fig.~\ref{fig:ESRFig3} (right frame). 
From matching the Curie-Weiss temperature at $x=0$, $\theta(x\!=\!0) = 11.2$\,K (taken from magnetic susceptibility), we can extract a nearest-neighbor coupling of $J_1=7.5$\,K if we assume $J_2=0$, or $J_1=6.7$\,K for $J_2/J_1=0.12$, the former in agreement with Ref.~\onlinecite{schmidt_yb_2021}.

For a quantitative comparison between the simulation and experiment of the full temperature dependence of $\chi$, we need to take into account that the natural energy scale for the classical magnet is $J_1 S^2$ while this is replaced for the quantum magnet by $J_1 S(S+1)$. The plot of the susceptibility data with the data from the simulation shows a fairly good agreement (Figure S12, \cite{Supplement}). The best match was found for $(J_1 S^2)_{\rm cl}=4$\,K, corresponding to an exchange coupling $J_1$ of about $5.3$\,K, but some deviations between calculation and experiment remain at low temperatures in particular for doped samples. They may originate in additional contributions to the experimental susceptbility, either intrinsic (Van Vleck) or extrinsic (background signal of the sample holder), but they may also reflect the fact that the $J_1$-$J_2$ Heisenberg model is not sufficient to fully capture the physics of the material which in fact can be expected to display significant spin-orbit coupling.\\


\section{Conclusion}
The solid-solution series {\nayblus} was established for $0 \le x \le 1$ where Yb and Lu occupy the \textit{B} site of the $3R$ $\alpha$-NaFeO$_2$ structure statistically. The lattice parameters of the series ideally follow Vegard’s rule and neither evidence for a phase segregation nor for an ordered superstructure was found. The band gaps change only for high Lu contents and the magnetic susceptibility reflects the amount of Yb$^{3+}$ ions. The ESR spectra of {\nayblus} show clear dependencies on the Yb$^{3+}$ content. It was shown that small differences in the lattice parameters upon changing the Yb$^{3+}$ amount produce small changes in the crystal fields and the $g$ values of the effective spin-$1/2$ Kramers Yb$^{3+}$ doublets. Also, the interaction among the Yb$^{3+}$ spins, as reflected in the ESR linewidth and ESR intensity, as well as in Weiss temperatures, changes with Yb$^{3+}$ amount. The line shape for the lowest investigated Yb amount of $x = 0.9$ has clear features typical for a hyperfine splitting, indicating single-ion behavior of independent Yb$^{3+}$ spins. 

In summary, we have shown that the magnetic exchange network of {\naybs} can be tuned by a random Lu substitution on the Yb site while maintaining the strong frustration. Further measurements, such as heat capacity, AC suceptibilty or $\mu$SR sepctroscopy, in the sub-Kelvin regime are necessary and scheduled to examine if the QSL state is preserved upon Lu doping up to the percolation threshold, as suggested by our present results. Thus, non-magnetic substitution opens up a new direction in QSL research, e.g., in order to elucidate to what extent the magnetic field--temperature phase diagram is affected by the controlled dilution.


\section{Acknowledgment}
The work has been financially supported by the DFG through SFB 1143 (project id 247310070) and the W\"urzburg-Dresden Cluster of Excellence on Complexity and Topology in Quantum Matter---\textit{ct.qmat} (EXC 2147, project id 390858490). ECA was supported by CNPq (Brazil) Grants No.\ 406399/2018-2 and No.\ 302994/2019-0, and FAPESP (Brazil) Grant No.\ 2019/17026-9.
We highly appreciate the magnetic evaluation of our materials by Heike Rave and the conduction of elemental analyses by Dr. Gudrun Auffermann (both MPI-CPfS Dresden).


\bibliographystyle{apsrev4-2} 
\bibliography{references}
\end{document}